\documentclass{PoS}

\newcommand{\amuhvp}{a_\mu^\textnormal{\scriptsize HVP}}%a_\mu^{hlbl}
\newcommand{\amulbl}{a_\mu^\textnormal{\scriptsize HLBL}}

\title{Recent Developments of Muon g-2 from Lattice QCD}

\ShortTitle{Recent Developments of Muon g-2 from Lattice QCD}

\author{\speaker{Vera G\"ulpers}\\%\thanks{A footnote may follow.}\\
        School of Physics and Astronomy, University of Edinburgh,
	Edinburgh EH9 3JZ, UK\\\\
        E-mail: \email{Vera.Guelpers@ed.ac.uk}}

\abstract{One of the most promising quantities for the search of signatures of physics beyond the 
Standard Model is the anomalous magnetic moment $g-2$ of the muon, where a comparison of the 
experimental result with the Standard Model estimate yields a deviation of about $3.5~\sigma$. On 
the theory side, the largest uncertainty arises from the hadronic sector, namely the hadronic 
vacuum polarisation and the hadronic light-by-light scattering. I review recent progress in 
calculating the hadronic contributions to the muon $g-2$ from the lattice and discuss the prospects 
and challenges to match the precision of the upcoming experiments.}

\FullConference{37th International Symposium on Lattice Field Theory - Lattice2019\\
		16-22 June 2019\\
		Wuhan, China}

\begin{document}

\section{Introduction}
The anomalous magnetic moment of the muon $a_\mu=(g_\mu-2)/2$ is one of the most promising 
quantities for 
the search of physics beyond the Standard Model of particle physics, since it can be measured 
and calculated to a high precision. %\par
The current most precise experimental determination of the anomalous magnetic moment of the muon 
has been obtained using polarised muons in a storage ring at 
Brookhaven National Laboratory \cite{Bennett:2006fi}
 \begin{equation}
   a^\textrm{\scriptsize exp}_\mu = 11659209.1 ( 5.4 )( 3.3 ) \times 10^{-10}\,.
   \label{eq:amuexp}
 \end{equation}
 Two upcoming experiments at Fermilab \cite{Venanzoni:2014ixa} and JPARC \cite{Otani:2015jra} plan 
to further reduce the uncertainty of the experimental result by a factor of four.
\par
\begin{figure}[h]
\centering
\includegraphics[scale=1.2]{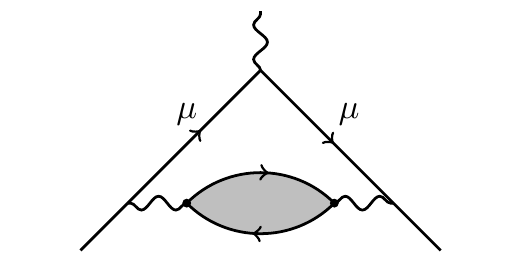}\hspace{1.5cm}
\includegraphics[scale=1.2]{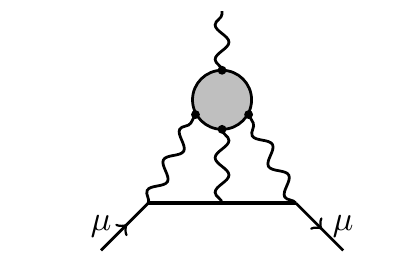}
\caption{Hadronic contributions to $a_\mu$: The hadronic vacuum polarisation (left) and the 
hadronic light-by-light scattering (right).}
\label{fig:hadrdiagrams}
\end{figure}

When estimating the  Standard Model prediction of $a_\mu$, contributions from the different 
fundamental 
interactions need to be calculated and a summary of the current most precise results is given in 
table \ref{tab:amucontr}. The biggest contribution to $a_\mu$ comes from the electromagnetic 
interaction (em) whereas the error is dominated by the QCD contributions, 
namely the hadronic vacuum polarisation (HVP) and the hadronic light-by-light scattering (HLBL). 
The corresponding diagrams for these processes are shown in figure \ref{fig:hadrdiagrams}.
\par
 \begin{table}[h]
 \centering
  \begin{tabular}{|l|r|r|}
  \hline
  & $a_\mu\times10^{10}$ & reference\\
  \hline\hline
  em & $11658471.895(8)$ & \cite{Aoyama:2012wk}\\
  weak & $15.36(10)$ & \cite{Gnendiger:2013pva}\\[0.3cm]
  HVP & $693.26(2.46)$ & \cite{Keshavarzi:2018mgv}\footnotemark\\
  HVP (NLO) & $-9.84(6)$ & \cite{Hagiwara:2011af}\\
  HVP (NNLO) & $1.24(1)$ & \cite{Kurz:2014wya} \\
  HLBL & $10.5(2.6)$ & \cite{Prades:2009tw} \\
  \hline
  total & $11659182.4(1)(2.5)^\textrm{\scriptsize HVP}(2.6)^\textrm{\scriptsize HLBL}$ & \\
  \hline
  \end{tabular}
  \caption{Standard Model Contributions to $a_\mu$}
  \label{tab:amucontr}
 \end{table}
 \footnotetext[1]{Other recent determinations of the HVP using $R$-ratio data can be found in 
\cite{Davier:2019can,Jegerlehner:2017lbd}.}
\par
A comparison of the Standard Model prediction of $a_\mu$ (cf. table \ref{tab:amucontr}) and the 
experimental determination reveals a deviation of about $3.5\sigma$, which could potentially hint 
to new physics. Further investigation requires both, the experimental as well as 
the theoretical uncertainties to be reduced. To match the targeted precision of the upcoming 
experiments 
($\approx1.5\times 10^{-10}$), this requires know\-ledge of the HVP contribution at the level of 
$0.2\%$ accuracy and the HLBL contribution at a level of about $10\%$.
\par
In recent years a lot of effort has been undertaken to calculate the hadronic contributions to the 
anomalous magnetic moment of the muon from first principles using Lattice QCD. In the remainder of 
this proceedings I will review the 
current status of such lattice calculations and discuss the prospects and challenges to match the 
precision of the upcoming experiments. 
\section{Hadronic Vacuum Polarisation}
Currently the most precise determinations of the HVP contribution to 
$a_\mu$ are obtained by using a dispersion relation and experimental data for 
the cross section $\sigma(e^+\,e^-\rightarrow
\textnormal{hadrons})$ as an input
\begin{equation}
\amuhvp = \left(\frac{\alpha 
 m_\mu}{3\pi}\right)^2
 \int\limits_{m^2_\pi}^\infty \textnormal{d} s\, \frac{R(s) K(s)}{s^2}
 \hspace{1.5cm}\textnormal{with}\qquad
 R(s)=\frac{\sigma(e^+\,e^-\rightarrow
\textnormal{hadrons},s)}{\sigma(e^+\,e^-\rightarrow 
\mu^+\mu^-,s)}
\end{equation}
with an analytically known kernel function $K(s)$. Recent determinations of $\amuhvp$ from this 
method can be found in 
\cite{Keshavarzi:2018mgv,Davier:2019can,Jegerlehner:2017lbd} and have a precision of about $0.5\%$. 
However, this approach relies on experimental input and an \textit{ab initio} calculation can be 
done using lattice QCD.

\subsection{The HVP from the Lattice}
\begin{figure}[h]
 \centering
 \includegraphics[width=0.7\textwidth]{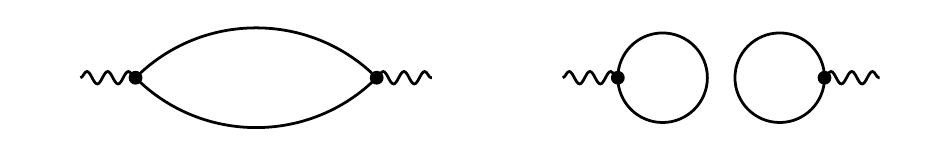}
 \vspace{-0.3cm}
 \caption{Quark-connected (left) and quark-disconnected (right) Wick contraction for the HVP.}
 \label{fig:Wick}
\end{figure}

In the following we will discuss how the HVP contribution to $a_\mu$ can be calculated using 
lattice QCD. The hadronic vacuum polarisation tensor   
\begin{equation}
 \Pi_{\mu\nu}(Q) \equiv 
\int\!\textnormal{d}^4 x\,\,
e^{i\,Q\cdot
 x}\,\left<j^\gamma_\mu(x)\,\,j^\gamma_\nu(0)\right>
 = (Q_\mu Q_\nu -
 \delta_{\mu\nu} Q^2)\,\Pi(Q^2)\,,
 \end{equation}
 is given by the four-dimensional Fourier transform of the 
correlation of two electromagnetic currents
\begin{equation}
 j^\gamma_\mu = \frac{2}{3}
\overline{u}\gamma_\mu u -  \frac{1}{3}
  \overline{d}\gamma_\mu d -  \frac{1}{3}
  \overline{s}\gamma_\mu s + \frac{2}{3}
\overline{c}\gamma_\mu c\,,
\end{equation}
which receive contributions from the different quark flavours multiplied by the respective charge 
factors. The contribution to the anomalous magnetic moment can then be determined from the vacuum 
polarisation function $\Pi(Q^2)$ as \cite{Blum:2002ii}
\begin{equation}
 \amuhvp = \left(\frac{\alpha}{\pi}\right)^2
\int\limits_0^\infty \textnormal{d} Q^2\, K(Q^2)\,\hat{\Pi}(Q^2)
\hspace{0.9cm}\textnormal{with}\hspace{0.4cm}\hat{\Pi}(Q^2) =
4\,\pi^2\left[\Pi(Q^2) - \Pi(0)\right]\,.
\label{eq:amusubHVP}
\end{equation}
with an analytically known electromagnetic kernel function $K(Q^2)$. In the last few years it has 
become common to calculate the subtracted vacuum polarisation $\hat{\Pi}(Q^2)$ required in 
equation~(\ref{eq:amusubHVP}) directly from the vector-vector two-point function $C(t)$ projected 
to zero spatial momentum \cite{Bernecker:2011gh,Feng:2013xsa}
\begin{equation}
\hat{\Pi}(Q^2) = 
4\pi^2\!\!\!\int\limits_0^\infty\!\!\textnormal{d} 
t\,C(t)\!\left[\!\frac{\cos(Qt)\!-\!1}{Q^2}+\frac{1}{2}t^2\right] 
\hspace{1cm}\textnormal{with}\qquad
C(t) = 
\frac{1}{3}\sum\limits_{k=0}^2 \sum\limits_{\vec{x}}
 \left<j^\gamma_k\!(\vec{x},t)j^\gamma_k\!(0)\right>\,.
\end{equation}
Changing the order of integration one can obtain $a_\mu$ directly from $C(t)$ by integrating over 
the Euclidean time using the appropriate kernel function $f(t)$ 
\begin{equation}
\hspace{0.6cm}\amuhvp\! = \!\!\!\int\limits_0^\infty\!\!\textnormal{d} t\,f(t) C(t) 
\,.
\end{equation}
In isospin symmetric QCD\footnote{We will discuss isospin breaking corrections in section 
\ref{subsec:IB}} the vector two-point function can be written in the following flavour decomposition
\begin{equation}
 C(t) = \frac{5}{9} C^\ell(t) + \frac{1}{9} C^s(t) + \frac{4}{9} C^c(t) + 
C^\textnormal{\scriptsize disc}(t)\,,
\end{equation}
with connected contributions for the light quark $C^\ell$, strange quark $C^s$ and charm $C^c$ 
quark as well as 
a quark-disconnected contribution $C^\textnormal{\scriptsize disc}$. The diagrams corresponding to 
the quark-connected and 
quark-disconnected Wick contractions are shown in figure \ref{fig:Wick}.
\par
In the following, I will discuss the contributions to $\amuhvp$ flavour by flavour, starting with 
the 
light-quark connected contribution in section~\ref{subsec:light}, followed by the strange- and 
charm-quark connected
contributions in section~\ref{subsec:strange}, quark-disconnected contributions in 
section~\ref{subsec:disc} and finally isospin breaking corrections in section~\ref{subsec:IB}. A 
summary and comparison of the various available results for $\amuhvp$ as 
well as an outlook for this quantity is given in section \ref{subsec:HVPsummary}.
\subsection{Light-Quark Contribution}
\label{subsec:light}
In this section, I will show results for the light-quark connected contribution to $\amuhvp$ and
discuss some of the main challenges for achieving a sub-percent precision calculation, namely the 
long distance 
noise-to-signal problem, finite volume corrections and accurate scale setting. 
\subsubsection{Long-Distance tail of the Vector Correlator}
\begin{figure}[h]
 \centering
 \includegraphics[width=0.39\textwidth]{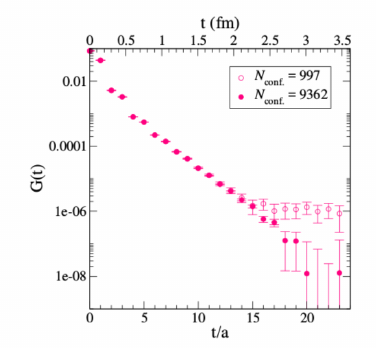}\hspace{1.3cm}
 \includegraphics[width=0.49\textwidth]{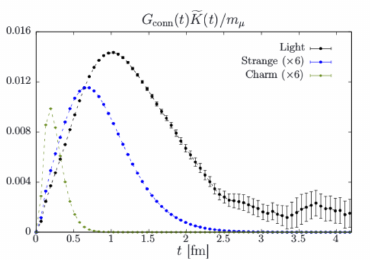}
 \caption{The vector correlator at physical pion mass: The plot on the left is from 
\cite{Davies:2019efs} and shows the two-point function $C(t)$ for two ensembles with the 
same parameters and different statistics. The plot on the right is from \cite{Gerardin:2019rua} and 
shows the integration kernel $f(t)\cdot C(t)$ for $\amuhvp$ with the light-quark contribution shown 
in black.}
\label{fig:lightcorrelator}
 \end{figure}

Figure \ref{fig:lightcorrelator} shows examples for the light-quark vector-vector two-point 
function $C(t)$
from the \linebreak HPQCD/Fermilab/MILC \cite{Davies:2019efs} collaboration (left) and the 
integration kernel $f(t)\cdot C(t)$ 
for $\amuhvp$ from Mainz \cite{Gerardin:2019rua} (right). Both data sets have been obtained at 
physical pion mass.
\par
As one can see in both plots, the signal in the vector two-point function deteriorates for large 
Euclidean times $t$. The statistical error on the raw data can be improved by using noise reduction 
techniques such as all-mode-averaging \cite{Blum:2012uh, Bali:2009hu} or low-mode-averaging, 
which has been successfully used in \cite{Blum:2018mom, Aubin:2019usy}.
However, the statistical uncertainty for $\amuhvp$ after integrating the raw 
data of $C(t)$ over $t$ will 
still be dominated by the noise from large times. A possible approach to reduce this uncertainty, 
is 
to replace the correlator by a (multi-) exponential fit for $t>t_c$ (see, e.g., 
\cite{DellaMorte:2017dyu, Davies:2019efs}). A more systematic way to treat the long distance tail 
of the correlator is the bounding 
method \cite{Borsanyi:2016lpl, Blum:2018mom}, which will be discussed in the following.
\subsubsection{Bounding Method}
\begin{figure}[h]
 \centering
 \includegraphics[width=0.98\textwidth]{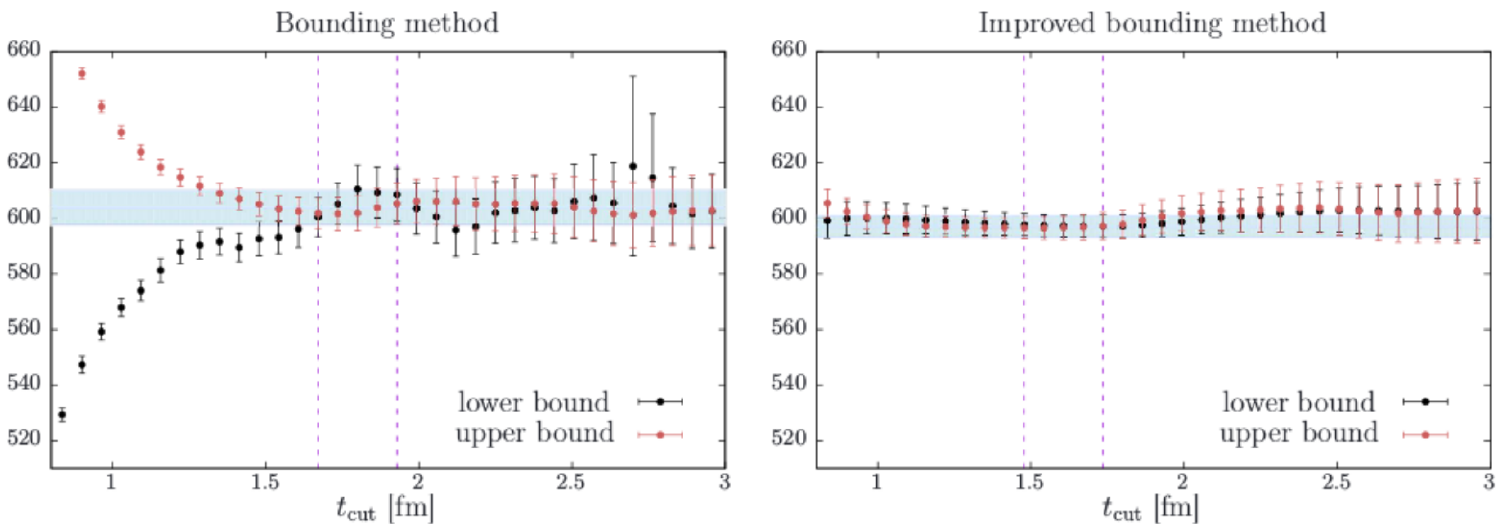}
 \caption{$\amuhvp$ from the bounding method (left) and the improved bounding 
method (right) plotted against $t_c$. The plots are taken from \cite{Gerardin:2019rua} and use a 
pion mass of $M_\pi\approx200$~MeV.}
\label{fig:bounding}
\end{figure}

The vector correlator can be written using the spectral representation as the sum of exponentials 
with positive prefactors
 \begin{equation}
  C(t) = \sum_n \frac{A^2_n}{2E_n} e^{-E_n t}\hspace{1.2cm}\textnormal{with}\qquad A_n^2>0\,.
  \label{eq:specrep}
 \end{equation}
The idea of the bounding method is to bound the two-point function $C(t)$ for Euclidean times 
larger 
than some value $t_c$ from 
above and below
 \cite{Borsanyi:2016lpl,Blum:2018mom}
  \begin{equation}
0< C(t_c)\, e^{-E_{t_c} (t-t_c)}  \leq C(t) \leq C(t_c)\, e^{-E_0(t-t_c)}
\hspace{1cm}\textrm{for}\qquad t\geq t_c 
\,.
 \end{equation}
 As a trivial lower bound one can use zero, since the correlator (cf. equation (\ref{eq:specrep})) 
is 
strictly positive. A more stringent lower bound for $t\geq t_c$ can be obtained by
using the effective mass $E_{t_c}$ at the given $t_c$. Since energy states with higher energies 
decay faster, the true correlator $C(t)$ is guaranteed to fall slower than 
$C(t_c)\, e^{-E_{t_c} (t-t_c)}$ for $t\geq t_c$. On the other hand, the true correlator is 
guaranteed to decay faster than the actual ground state energy $E_0$, and thus $C(t)\leq C(t_c)\, 
e^{-E_0(t-t_c)}$ is an upper bound for the correlator. In the vector channel, the ground state 
energy $E_0$ is given by the finite volume energy of two pions with one unit of momentum. When 
calculating $\amuhvp$ one now replaces the correlator for $t\geq t_c$ by the upper and the lower 
bound and varies $t_c$. An example of $\amuhvp$ plotted against $t_c$ is shown on the left-hand 
side of figure \ref{fig:bounding}. As one can see, once $t_c$ is large enough, the result 
using the upper and lower bound overlap, and $\amuhvp$ can be extracted from this region. 
\par

\par
\begin{figure}[h]
  \centering
  \includegraphics[width=0.48\textwidth]{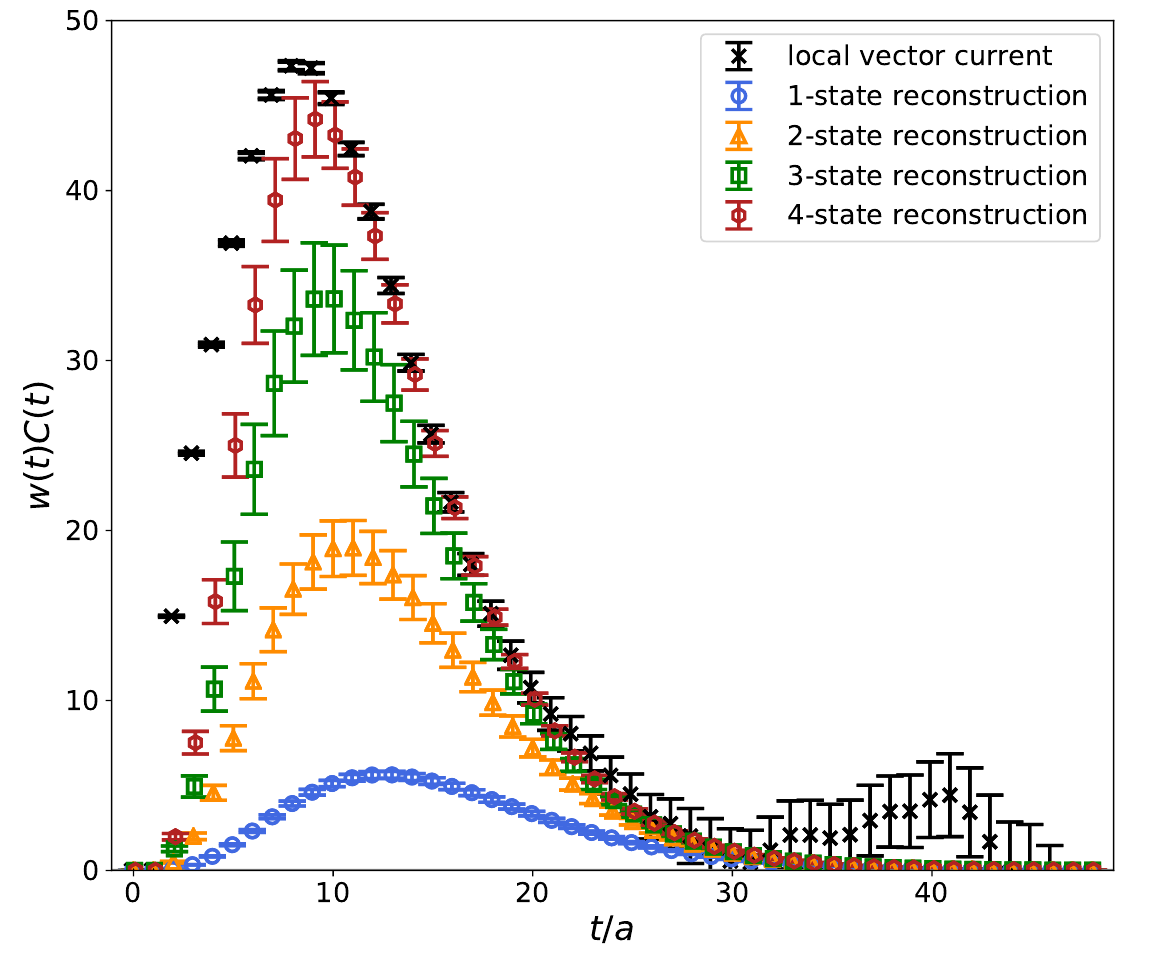}
  \caption{The integration kernel $f(t)\cdot C(t)$ for $\amuhvp$ plotted against $t$. The black 
data show the original correlator, the coloured points are reconstructions of the vector two-point 
function using one (blue) and up to four (red) states from the GEVP. The figure is taken from 
\cite{Bruno:2019nzm}.}
\label{fig:corrreconstruction}
 \end{figure}

The bounding method can be further improved \cite{Gerardin:2019rua,Bruno:2019nzm} if a dedicated 
spectroscopy study for the vector channel is available. This requires to calculate two-point 
functions using various operators that have overlap with two-pion states. With the 
Generalised Eigenvalue Problem (GEVP)~\cite{Blossier:2009kd}, one can extract the energies 
$E_n$ and overlap factors $A_n$ for the lowest $N$ states of the spectrum.
Once these energies and overlap factors
have been determined, the long-distance tail of the vector correlator can be reconstructed. An 
example at physical pion mass is shown in figure \ref{fig:corrreconstruction}. As one can see, the 
more states are used in the reconstruction, the closer the reconstructed data are to the original 
correlator (shown in black in figure \ref{fig:corrreconstruction}). 
 \par
These reconstructed states can now be used to improve the bounding method as follows. Subtract the 
lowest $N$ states from the correlation function and bound the subtracted correlator $\tilde{C}(t)$  
 \begin{equation}
  \tilde{C}(t) = C(t) - \sum_{n=0}^{N-1} \frac{A^2_n}{2E_n} e^{-E_n t}
\hspace{1.9cm}
0\leq \tilde{C}(t_c)\, e^{-E_{t_c} (t-t_c)}  \leq \tilde{C}(t) \leq \tilde{C}(t_c)\, 
e^{-E_N(t-t_c)}\,.
 \end{equation}
 Here, the upper bound is obtained using the energy of the $(N+1)$th state, i.e.\ the lightest 
state that was not subtracted from the correlator. The right-hand side of figure \ref{fig:bounding} 
shows the improved bounding method from \cite{Gerardin:2019rua} with the two lightest states 
subtracted. As one 
can see in comparison with the unimproved bounding method (left on figure \ref{fig:bounding}) the 
upper and the lower bound now overlap at much smaller $t_c$ and $\amuhvp$ can be extracted with a 
smaller error. 

\subsubsection{Finite Volume Effects}
Finite volume (FV) effects in the vector correlator are dominated by the two-pion state and can 
thus be expected to be important for large time separations $t$. %\par
Various studies (see, e.g.~\cite{Shintani:2019wai,Aubin:2019usy,Gerardin:2019rua,LehnerLat19}) of 
FV effects for lattice calculations of $\amuhvp$ suggest that these are of the order of 
$\Delta^{\textnormal{\scriptsize FV}}\amuhvp \approx 20-30\times10^{-10}$ for typical sizes 
of $5-6$~fm of state-of-the-art lattice ensembles used at the 
physical point. Thus, it is crucial to carefully study and correct for FV effects when aiming at 
percent level precision for the HVP.
\par
\begin{figure}[h]
\centering
 \includegraphics[width=0.75\textwidth]{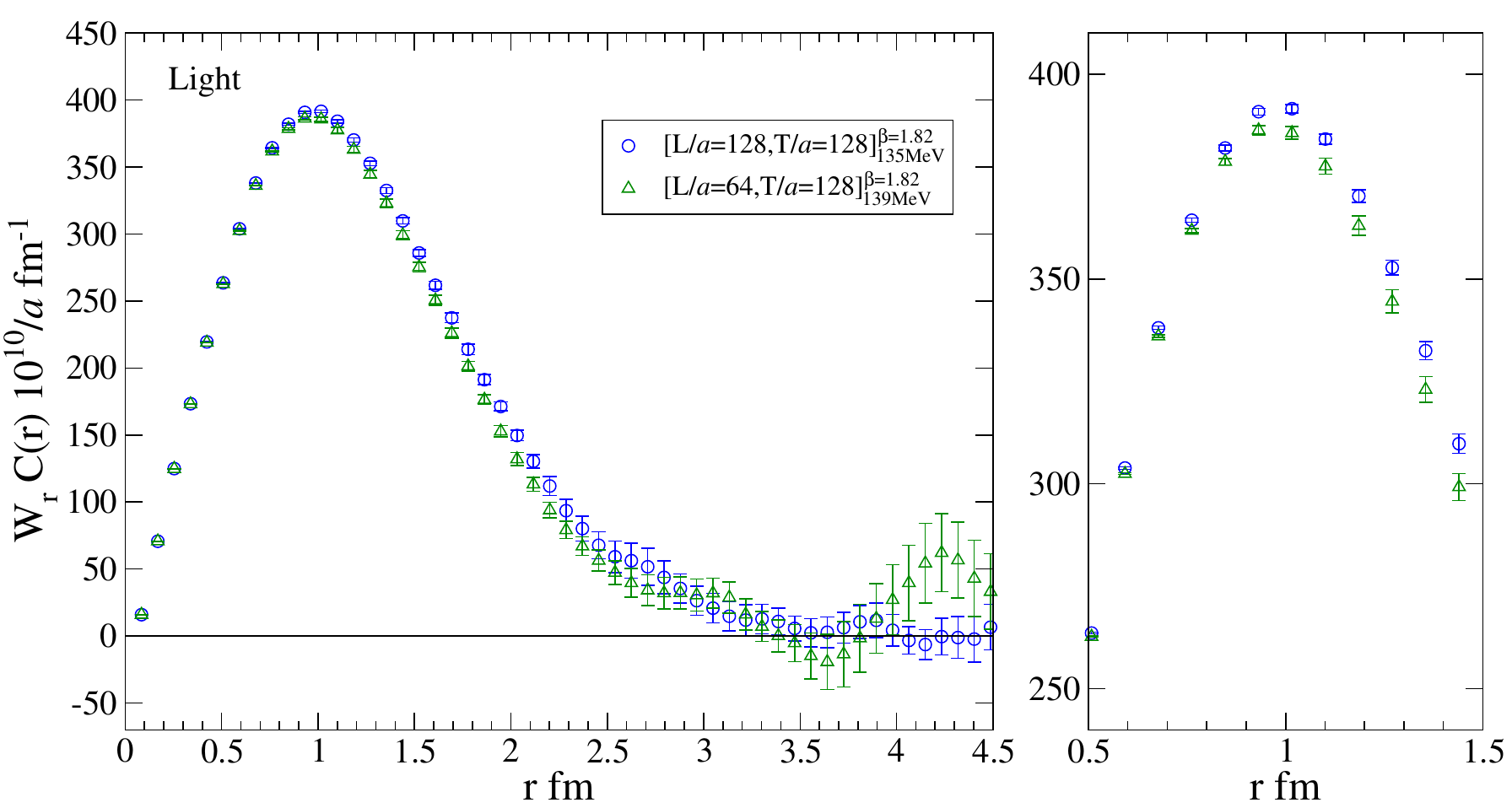}
 \caption{The $\amuhvp$ integrand calculated using two different volumes of $5.4$~fm (green 
triangles) and $10.8$~fm (blue circles) at physical pion mass. The plot on the right shows a 
zoomed-in version of the data on the left. The plot is taken from \cite{Shintani:2019wai}.}
\label{fig:FVShintanietal}
\end{figure}

A straightforward way to study finite volume effects is using ensembles that 
differ only in the volume. Figure \ref{fig:FVShintanietal} shows results from the PACS 
collaboration 
\cite{Shintani:2019wai} for the $\amuhvp$ integrand calculated using two different volumes of 
$5.4$~fm and $10.8$~fm at the physical pion mass. One can clearly see, a significant difference 
between the data on both ensembles, in particular for larger values of $t$. The authors of 
\cite{Shintani:2019wai} found finite volume effects for $\amuhvp$ to be about $1.7$ times larger 
than what was expected from next-to-leading order (NLO) Chiral Perturbation Theory ($\chi$PT). 
\par
Finite volume corrections to $\amuhvp$ have been determined at next-to-next-to-leading order 
\linebreak (NNLO) 
in $\chi$PT recently \cite{Bijnens:2017esv,Aubin:2019usy} and the authors of \cite{Aubin:2019usy} 
find that additional FV effects from NNLO are about $0.4-0.45$ times the NLO FV corrections.
\par
\begin{figure}[h]
 \centering
 \hspace{-0.3cm}
 \includegraphics[width=0.52\textwidth,trim = 0mm 0mm 0mm 0mm, 
clip=true]{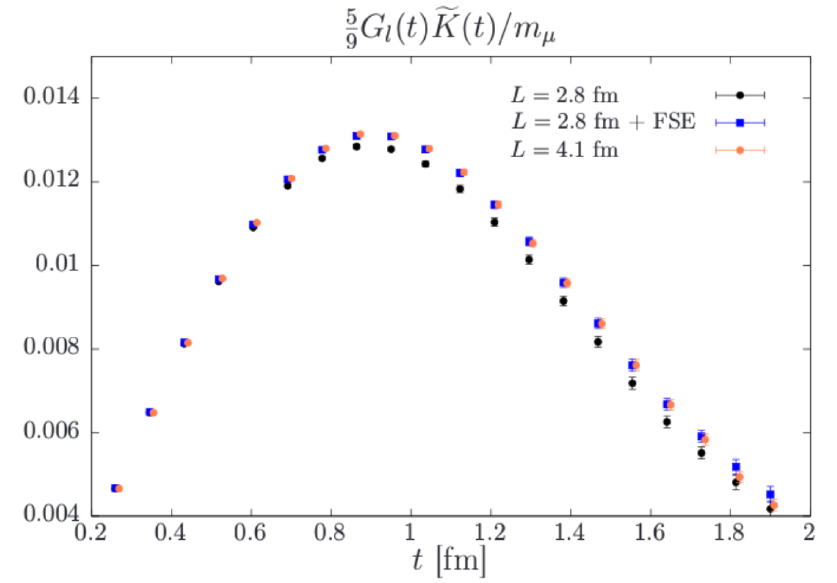}
 \hspace{0.7cm}
 \includegraphics[width=0.43\textwidth]{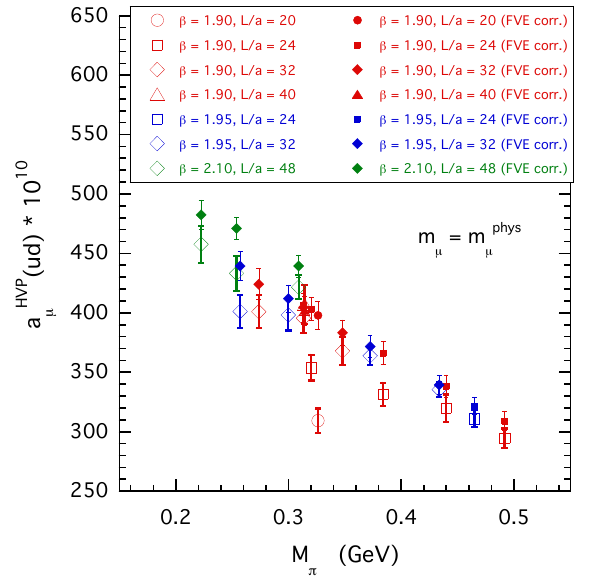}
 \caption{Correcting finite volume effects with the timelike pion form factor. The plot on the 
left is taken from \cite{Gerardin:2019rua} and shows the $\amuhvp$ integrand for two different 
ensembles at the same pion mass $M_\pi=280$~MeV with different volumes. The smaller volume is shown 
without (black circles) and with FV correction (blue squares). The plot on the right is taken from 
\cite{Giusti:2018mdh} and shows $\amuhvp$ vs the pion mass for various ensembles without (open 
symbols) and with (closed symbols) FV corrections.}
 \label{fig:FVtimelikepff}
\end{figure}

A systematic way to study and correct for FV effects for the HVP is by writing the long 
distance contribution of the vector two-point function in terms of the timelike pion form factor.
In \cite{Meyer:2011um, Francis:2013qna} it was suggested to use the Gounaris-Sakurai 
parameterization of the timelike pion form factor to calculate the infinite volume long-distance 
vector two-point function and the finite volume equivalent using the Gounaris-Sakurai 
parameterization combined with the Lellouch-L\"uscher formalism 
\cite{Luscher:1991cf,Lellouch:2000pv}. This approach has been used by Mainz 
\cite{Gerardin:2019rua}, ETMC \cite{Giusti:2018mdh} and RBC/UKQCD \cite{LehnerLat19} to study 
FV volume corrections and results after correcting for FV effects are found to be consistent when 
comparing 
ensembles that only differ by volume \cite{Gerardin:2019rua,Giusti:2018mdh}. The plot on the 
left-hand side of figure \ref{fig:FVtimelikepff} is from Mainz \cite{Gerardin:2019rua} and shows the 
$\amuhvp$ integrand for two different 
ensembles at the same pion mass $M_\pi=280$~MeV with different volumes. The smaller volume is shown 
without (black circles) and with (blue squares) FV correction using the timelike pion form factor. 
After the data on the smaller ensemble has been corrected the results for the two different volumes 
agree with each other. The plot on the right-hand side of figure \ref{fig:FVtimelikepff} is from 
ETMC \cite{Giusti:2018mdh} and shows $\amuhvp$ plotted against the pion mass for various ensembles 
without (open symbols) and with (closed symbols) FV corrections. In addition to using the timelike 
pion form factor to estimate FV effects for large times $t$, the 
authors of \cite{Giusti:2018mdh} use perturbative QCD for small $t$ to correct for FV effects. At a
pion mass of around $M_\pi=320$~MeV, where several ensembles are available that only differ in 
volume, results are found to be in agreement once FV effects have been corrected for.

\par
\par
In a recent paper \cite{Hansen:2019rbh}, FV corrections to the HVP have been studied using a 
Hamiltonian approach, currently quoting corrections at $\mathcal{O}(e^{-m_\pi L})$, but neglecting 
effects of $\mathcal{O}(e^{-\sqrt{2}m_\pi L})$ and higher orders.

\subsubsection{Scale Setting}
Although $a_\mu$ is a dimensionless quantity, it depends on the scale of a given lattice, since the 
evaluation of the Kernel function for integrating the HVP requires to input the muon mass in 
lattice units. Assuming the scale has been set by some quantity $\Lambda$ with statistical error 
$\Delta\Lambda$, using error propagation one can show \cite{DellaMorte:2017dyu}, that the relative 
error on $\Lambda$ is going to be enhanced by a factor of $\approx 1.8$ for $\amuhvp$
 \begin{equation}
  \Delta \amuhvp = \left|\Lambda\,\, \frac{\textrm{d} \amuhvp}{\textrm{d} 
\Lambda}\right|\cdot\,\frac{\Delta \Lambda}{\Lambda} = \left|M_\mu\frac{\textrm{d} 
\amuhvp}{\textrm{d} 
M_\mu}\right|\cdot\frac{\Delta \Lambda}{\Lambda} \hspace{2cm}\textnormal{with}\qquad M_\mu = 
\frac{m_\mu}{\Lambda}\,.
 \end{equation}
Thus, achieving $0.2\%$ accuracy on the HVP requires knowledge of the lattice scale to at least 
$0.1\%$. A suitable quantity for high-precision scale setting might be the mass of the 
$\Omega$-Baryon. However, whether or not scale setting at $0.1\%$ accuracy is possible in the near 
future remains to be investigated.

\subsubsection{Comparison of Results for the Light-Quark Contribution}
\label{subsubsec:lightres}
\begin{figure}[h]
 \centering
 \includegraphics[width=1\textwidth]{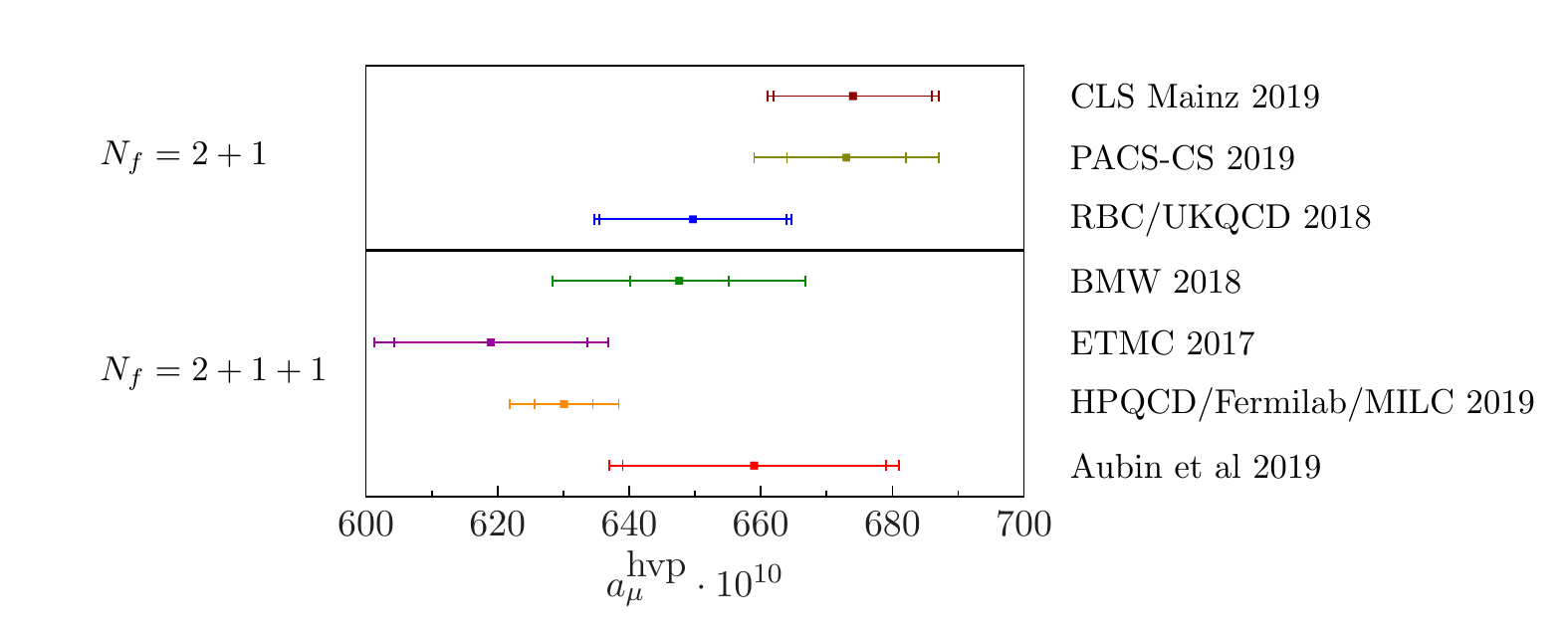}
 \vspace{-0.9cm}
 \caption{Comparison of lattice results for the light-quark contribution to $\amuhvp$. The values 
are 
taken from CLS Mainz 2019 \cite{Gerardin:2019rua}, PACS-CS 2019 \cite{Shintani:2019wai}, RBC/UKQCD 
2018 \cite{Blum:2018mom}, BMW 2018 \cite{Borsanyi:2017zdw}, ETMC 2018 \cite{Giusti:2018mdh}, 
HPQCD/Fermilab/MILC 2019 \cite{Davies:2019efs}, Aubin et al 2019 \cite{Aubin:2019usy}.}
 \label{fig:complight}
\end{figure}

Figure \ref{fig:complight} shows a comparison of results for the light-quark connected contribution 
to the 
HVP from various collaborations. The two different panels $N_f=2+1$ and $N_f=2+1+1$ denote the 
number of dynamical fermions used in the calculations. 
\par
\par
The relative errors on the light-quark contributions of the different results are about $1.3\%$ to 
$3.3\%$. For most of the collaborations, the error is dominated by statistics (inner error bar on 
the points in figure \ref{fig:complight}). As discussed above, the main challenge in terms of 
statistical error is the 
growing noise-to-signal ratio for large Euclidean time separations. Thus, a reduction of this error 
requires good control of the long-distance tail. A very promising approach for reaching sub-percent 
precision on the light quark contribution in the future is the improved bounding method discussed 
above. \par
As one can see, there is a slight tension between the smallest and the largest results of 
about $2\sigma$. This tension has to be object of further investigations, and in particular needs to 
be monitored when the collaborations further reduce the uncertainties in the individual 
calculations. A possible approach for 
determining the source of potential differences between different groups would be the comparison of 
more 
intermediate results, e.g. time-moments \cite{Chakraborty:2014mwa} of the vector correlator or 
$a_\mu$ calculated from a time window \cite{Blum:2018mom}.

\subsection{Strange- and Charm-Quark Contribution}
\label{subsec:strange}
The connected strange- and charm-quark contributions to the HVP are significantly 
smaller than the light-quark contribution and suffer far less from noise-to-signal problems in 
the long-distance tail or finite volume corrections. For the charm-quark contribution 
discretization effects can 
be large and lattice calculations should ideally include at least 
three different lattice spacings to reliably extrapolate to the continuum. 
\par
\begin{figure}[h]
 \centering
 \hspace{-0.5cm}
 \includegraphics[scale=0.73]{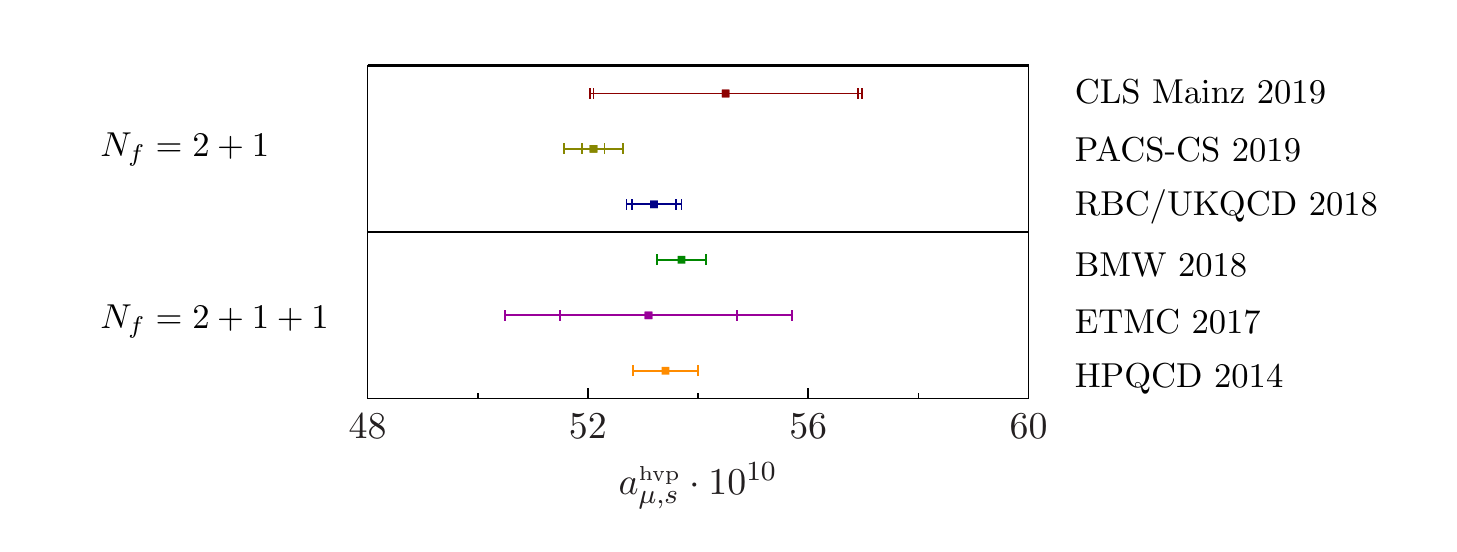}
\hspace{-0.05\textwidth}
\includegraphics[scale=0.73,trim = 36mm 0mm 42mm 0mm, 
clip=true]{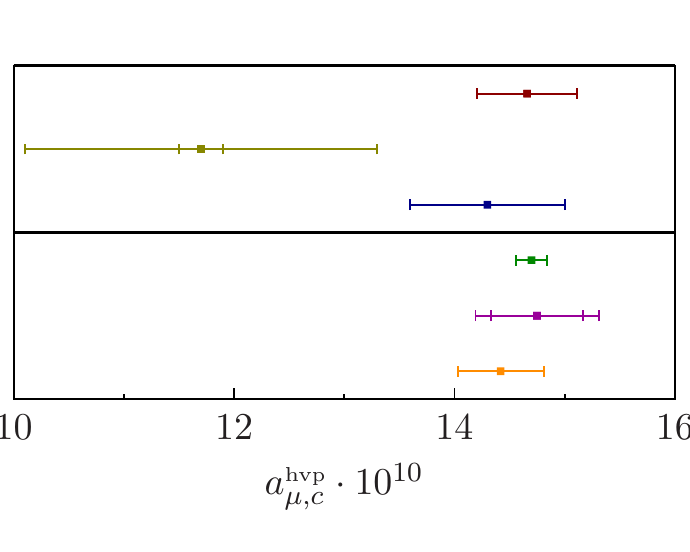}
\vspace{-0.4cm}
 \caption{Comparison of lattice results for the strange-quark (left) and charm-quark (right) 
contributions to $\amuhvp$. Results are taken from CLS Mainz 2019 \cite{Gerardin:2019rua}, PACS-CS 
2019 \cite{Shintani:2019wai}, RBC/UKQCD 
2018 \cite{Blum:2018mom}, BMW 2018 \cite{Borsanyi:2017zdw}, ETMC 2017 \cite{Giusti:2017jof}, 
HPQCD/Fermilab/MILC 2014 \cite{Chakraborty:2014mwa}.}
\label{fig:compsc}
\end{figure}

The results for the connected strange- and charm-quark contribution to $\amuhvp$ from various 
collaborations are shown in figure 
\ref{fig:compsc} and are in good agreement with each other. The errors correspond to errors of 
about $\lesssim0.4\%$ and $\lesssim0.3\%$ on the total HVP for the strange and charm, respectively. 
Thus, these contributions are already in a good shape when aiming at sub-percent precision for 
$\amuhvp$ from a lattice calculation.
\par
\par

\subsection{Quark-Disconnected Contribution}
\label{subsec:disc}
Besides the quark-connected contributions discussed above, the HVP receives a contribution from a 
quark-disconnected Wick contraction (cf. right-hand side of figure \ref{fig:Wick}). The calculation 
of the respective 
disconnected quark-loops requires knowledge of the propagator from all lattice points to all other 
lattice points (all-to-all propagator), which has to be determined stochastically and is thus 
notoriously noisy. 
The combined light- and 
strange-quark disconnected contribution is $SU(3)$-flavour suppressed, i.e.\ it would vanish if 
$m_s=m_\ell$. 
In \cite{Francis:2014hoa} it was shown that a substantial reduction in the statistical error can be 
achieved by exactly implementing $SU(3)$ suppression in the lattice calculation
\begin{equation}
 C^\textnormal{disc}(t) = \frac{1}{9} 
\left<(\Delta^\ell(t)-\Delta^s(t))\cdot(\Delta^\ell(0)-\Delta^s(0))\right>\hspace{1cm}\textrm{with}
\hspace{0.5cm} \Delta_\mu^f(t) = \sum_{\vec{x}} \textnormal{Tr}\left[\gamma_\mu S^f(x,x)\right]
\end{equation}
and stochastically estimate the difference $\Delta^\ell(t)-\Delta^s(t)$ rather than the individual 
quark loops. Various stochastic estimators for $\Delta^\ell(t)-\Delta^s(t)$ and further 
noise reduction techniques have been proposed, e.g.\ low-mode-averaging and sparsened noise sources 
\cite{Blum:2015you}, hierarchical probing \cite{Stathopoulos:2013aci, Gerardin:2019rua} or 
frequency-splitting estimators \cite{Giusti:2019kff,HarrisLat19}.
\par
\begin{figure}[h]
 \centering
 \includegraphics[width=0.80\textwidth]{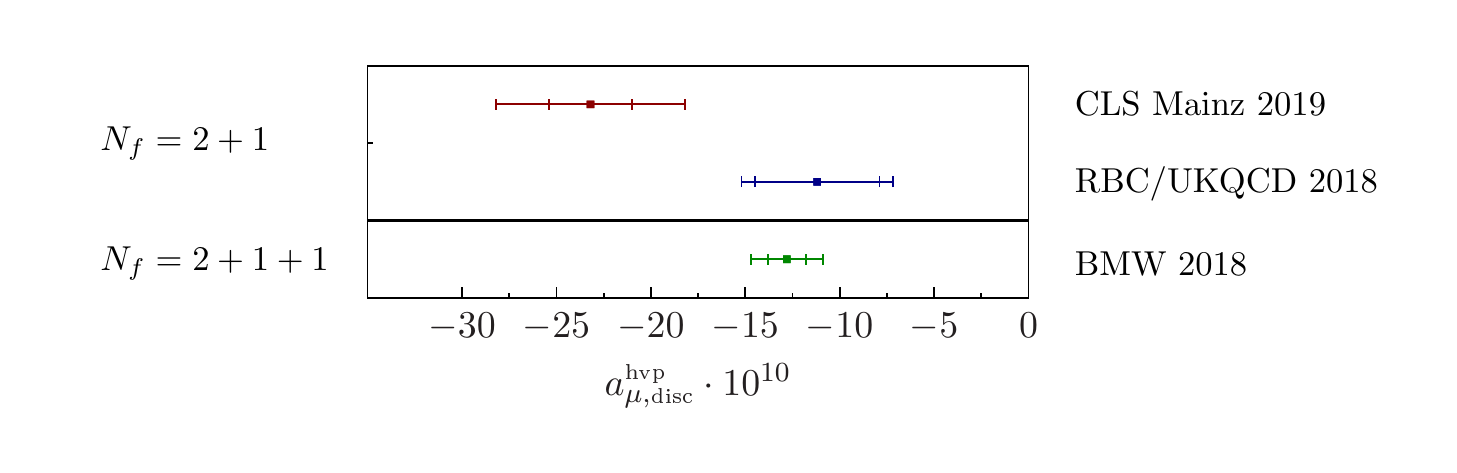}
 \vspace{-0.7cm}
 \caption{Comparison of lattice results for the quark-disconnected contribution to $\amuhvp$. 
Values 
are taken from CLS Mainz 2019 \cite{Gerardin:2019rua}, RBC/UKQCD 2018 \cite{Blum:2018mom}, BMW 2018 
\cite{Borsanyi:2017zdw}.}
\label{fig:disccomp}
 \end{figure}
Figure \ref{fig:disccomp} shows the published results for the quark-disconnected contribution to 
$\amuhvp$. The results are in reasonable agreement, although the results from Mainz is about 
$2\sigma$ below the two other values. Whether this is due to the chiral extrapolation done in 
\cite{Gerardin:2019rua} needs to be investigated by adding data closer or at the physical point in 
this calculation.
\par
The errors on the available published results for the quark-disconnected contribution correspond to 
errors of about 
$0.3-0.7\%$ on the total HVP  and thus, are already precise enough for a $1\%$ determination of 
$\amuhvp$, however, still need to be improved if an error of $<0.2\%$ is targeted. 
\par
Work in progress on calculating the quark-disconnected contribution to the hadronic vacuum 
polarisation 
by the HPQCD/Fermilab/MILC collaboration was presented at this conference 
\cite{Davies:2019acq}.

\subsection{Isospin Breaking Corrections}
\label{subsec:IB}
All lattice calculations discussed above have been done in the isospin symmetric limit, where the 
up- and the down-quark are treated as being equal. However, in nature, isospin is broken by the 
different electromagnetic charges of the up- and the down-quark as well as their different bare 
quark 
masses. These effects are expected to be of the order of $\mathcal{O}(\alpha)\approx1\%$ and 
$\mathcal{O}((m_d-m_u)/\Lambda_\textnormal{\scriptsize QCD})\approx1\%$, respectively. Clearly, a 
lattice 
calculation 
aiming at such precision will need to include these effects. It is important to stress 
that the separation of strong isospin breaking and QED effects require to define a renormalisation 
prescription. At the same time the definition of the ``physical'' point in a pure QCD simulation 
becomes scheme dependent. 
Only in full QCD$+$QED with $m_u\neq m_d$ the physical point is unambiguously 
defined, e.g.\ by matching a set of hadron masses to their experimental value (see e.g.\ 
\cite{Aoki:2019cca}).
\subsubsection{Strong Isospin Breaking Correction}
The effect from strong isospin breaking can be taken into account in a lattice calculation by 
simply using different input quark masses for the up and the down quark. This has been done for 
calculating the strong isospin breaking correction to the HVP by the HPQCD/Fermilab/MILC 
collaboration \cite{Chakraborty:2017tqp}. Here, 
two different gauge ensembles where used, one with $N_f=2+1+1$ and $m_u\neq m_d$ only in the 
valance sector and one with $N_f=1+1+1+1$ taking also strong isospin breaking corrections for the 
sea quarks into account. The strong isospin breaking correction is then quoted as the difference 
between calculating $\amuhvp$ using the average up- and down-quark mass for the light quarks and 
using 
the up-quark mass for the up and the down-quark mass for the down quark. In 
\cite{Chakraborty:2017tqp} the authors find $\delta^{sIB}\amuhvp = (7.7\pm3.7)\times10^{-10}$ and 
$\delta^{sIB}\amuhvp = (9.0\pm2.3)\times10^{-10}$ using the ensemble without and with $m_u\neq m_d$ 
for the sea quarks, respectively.
\par
\begin{figure}[h]
 \centering
 \includegraphics[width=0.99\textwidth]{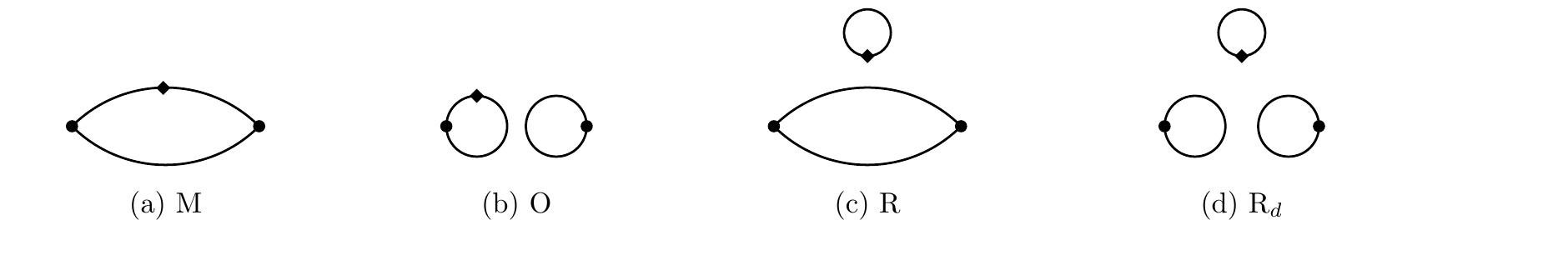}
 \vspace{-0.3cm}
 \caption{Mass insertion diagrams for the hadronic vacuum polarisation. The insertion of a scalar 
current is denoted by the diamond-shaped vertex.}
 \label{fig:sIBdiagrams}
\end{figure}

A different approach for including strong isospin breaking corrections in a lattice 
calculation is by expanding \cite{deDivitiis:2011eh} the path integral in the difference of the 
respective quark masses and 
their isospin symmetric mass $\hat{m}$
\begin{equation}
 \left<O\right>_{m_f\neq \hat{m}_f} = \left<O\right>_{m_f=\hat{m}} + 
\Delta m_f \left.\frac{\partial}{\partial 
m_f}\left<O\right>\right|_{m_f=\hat{m}} + \mathcal{O}\left(\Delta 
m_f^2\right)%\,,
\label{eq:patintegralmexpansion}
\end{equation}
with $\Delta m_f = m_f - \hat{m}$. At $\mathcal{O}(\Delta m_f)$ one has to calculate contributions 
with one 
insertion of a scalar current. The corresponding diagrams for the hadronic vacuum polarisation are 
shown in figure 
\ref{fig:sIBdiagrams}. Diagram $M$ and $O$ are the correction for the valance quarks for the 
quark-connected and quark-disconnected HVP, respectively, whereas diagrams $R$ and $R_d$ 
correspond to sea-quark effects. 
\par
The expansion of the path integral was used in 
\cite{Giusti:2019xct,Blum:2018mom} to calculate the valance effect for the quark-connected HVP 
(diagram $M$). The ETMC collaboration finds $\delta^{sIB}\amuhvp = (6.0\pm2.3)\times10^{-10}$ 
\cite{Giusti:2019xct} and RBC/UKQCD obtains $\delta^{sIB}\amuhvp = 
(10.6\pm4.3\pm6.6)\times10^{-10}$ \cite{Blum:2018mom}.

\subsubsection{QED Correction}
The determination of electromagnetic corrections requires the inclusion of QED when evaluating the 
Euclidean 
path integral. Since QED is a long range interaction, finite volume (FV) effects for lattice 
calculations can be large. Compared to pure QCD, where FV corrections are exponentially suppressed, 
in the case of QED finite volume corrections are usually power-law\footnote{A prescription for 
infinite volume 
QED without power-law finite volume corrections has been recently proposed in \cite{Feng:2018qpx}.}.
For QED$_L$ \cite{Hayakawa:2008an}, which is a commonly used prescription of QED in a finite volume, 
 all the spatial 
zero modes of the photon field are subtracted and finite volume effects for the QED 
corrections to the HVP are of $\mathcal{O}(1/(m_\pi L)^3)$ 
\cite{Bijnens:2019ejw,Hermansson-Truedsson:2019pna,Giusti:2017jof}, and thus negligible within the 
required precision for the HVP when using typical lattice sizes with $m_\pi L\gtrsim 4$. 
\par
\begin{figure}[h]
\centering
 \includegraphics[width=0.99\textwidth]{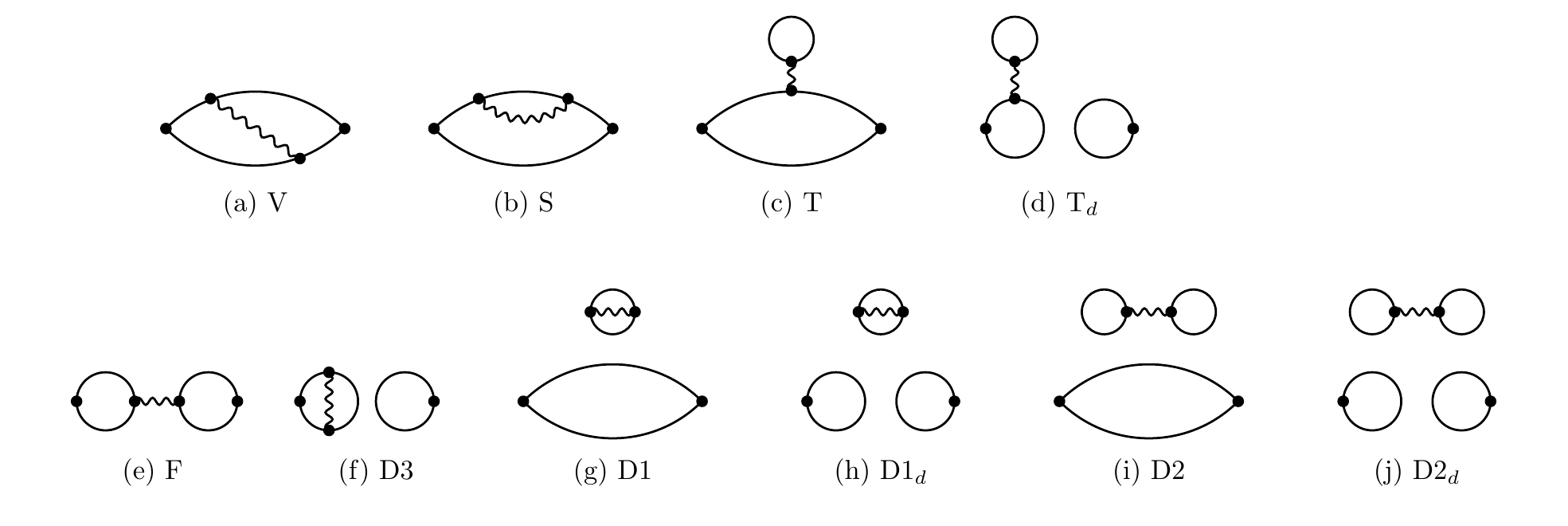}
\vspace{-0.3cm}
 \caption{QED correction diagrams for the hadronic vacuum polarisation.}
\label{fig:QEDdiagrams}
 \end{figure}

Since the electromagnetic fine structure constant $\alpha$ is small at low 
energies, QED can be treated in a perturbative way by expanding the QCD$+$QED path integral in 
$e^2$ \cite{deDivitiis:2013xla}
\begin{equation}
 \left<O\right>_{\textnormal{\scriptsize QCD}+\textnormal{\scriptsize QED}} = 
\left<O\right>_{\textnormal{\scriptsize QCD}} + 
\frac{1}{2}\,e^2\left.\frac{\partial^2}{\partial 
e^2}\left<O\right>\right|_{e=0} + {\cal{O}}(\alpha^2)\,.
\label{eq:pathintegraleexpansion}
\end{equation}
At $\mathcal{O}(\alpha)$ this requires to calculate diagrams that include one photon propagator. 
The respective diagrams for the HVP are shown in figure \ref{fig:QEDdiagrams}. Diagrams $S$ and $V$ 
are QED corrections to the quark-connected HVP and diagrams $F$ and $D3$ are QED corrections to the 
quark-disconnected HVP. All other diagrams correspond to QED effects for the sea quarks.
\par
The ETMC \cite{Giusti:2019xct} and RBC/UKQCD collaborations \cite{Blum:2018mom,Gulpers:2018mim} both 
have 
calculated QED corrections to the quark-connected HVP in the electro-quenched approximation, where 
QED corrections for the sea-quarks are not taken into account. Calculating QED corrections at 
various masses heavier than physical pion mass and extrapolating to the physical point, ETMC finds 
$\delta^\textnormal{\scriptsize QED} \amuhvp=1.1(1.0)\times 10^{-10}$ 
\cite{Giusti:2019xct}. RBC/UKQCD finds $\delta^\textnormal{\scriptsize QED} 
\amuhvp=5.9(5.7)\times 10^{-10}$ \cite{Blum:2018mom}, consistent with ETMC albeit 
with larger error bars, working directly at physical masses at a single lattice spacing. Further 
work in progress for QED corrections to the HVP by other collaborations was presented at this 
conference \cite{Risch:2019xio, TothLat19}.
\par
The leading QED correction to the quark-disconnected HVP is given by the diagram $F$ shown in 
figure \ref{fig:QEDdiagrams}. Other than the same diagram without the photon connecting the two 
quark loops (i.e.\ the pure QCD quark-disconnected diagram), this contribution is not 
$SU(3)$-flavour suppressed and could thus be important. When calculating this contribution, one is 
only interested in the case where, besides the photon line, the quark-loops are in addition 
connected by gluons. If no additional gluons connect the quarks, these contributions are 
conventionally included in the NLO HVP contribution\footnote{See \cite{Chakraborty:2018iyb} for a 
lattice calculation of the NLO HVP contribution to $a_\mu$} to $a_\mu$ (cf. table 
\ref{tab:amucontr}) and 
thus, need to be subtracted in a lattice calculation to avoid double counting.
RBC/UKQCD calculated this diagram and finds $\delta^{\textnormal{\scriptsize QED, disc}} 
\amuhvp =  -6.9(2.1)(1.4)\times 10^{-10}$ \cite{Blum:2018mom}. 
\par
The diagrams in figure \ref{fig:QEDdiagrams} corresponding to sea-quark effects are all at least 
either $SU(3)$-flavour or $1/N_c$ suppressed. However, naively, they could still be of the 
order of $\approx 33\%$ of the connected contribution. Thus, when aiming at sub-percent precision 
for the total HVP contribution to $a_\mu$, these effects will have to be included eventually.

\subsection{Summary - HVP Contribution to $a_\mu$}
\label{subsec:HVPsummary}
\begin{figure}[h]
 \centering
 \includegraphics[width=0.95\textwidth]{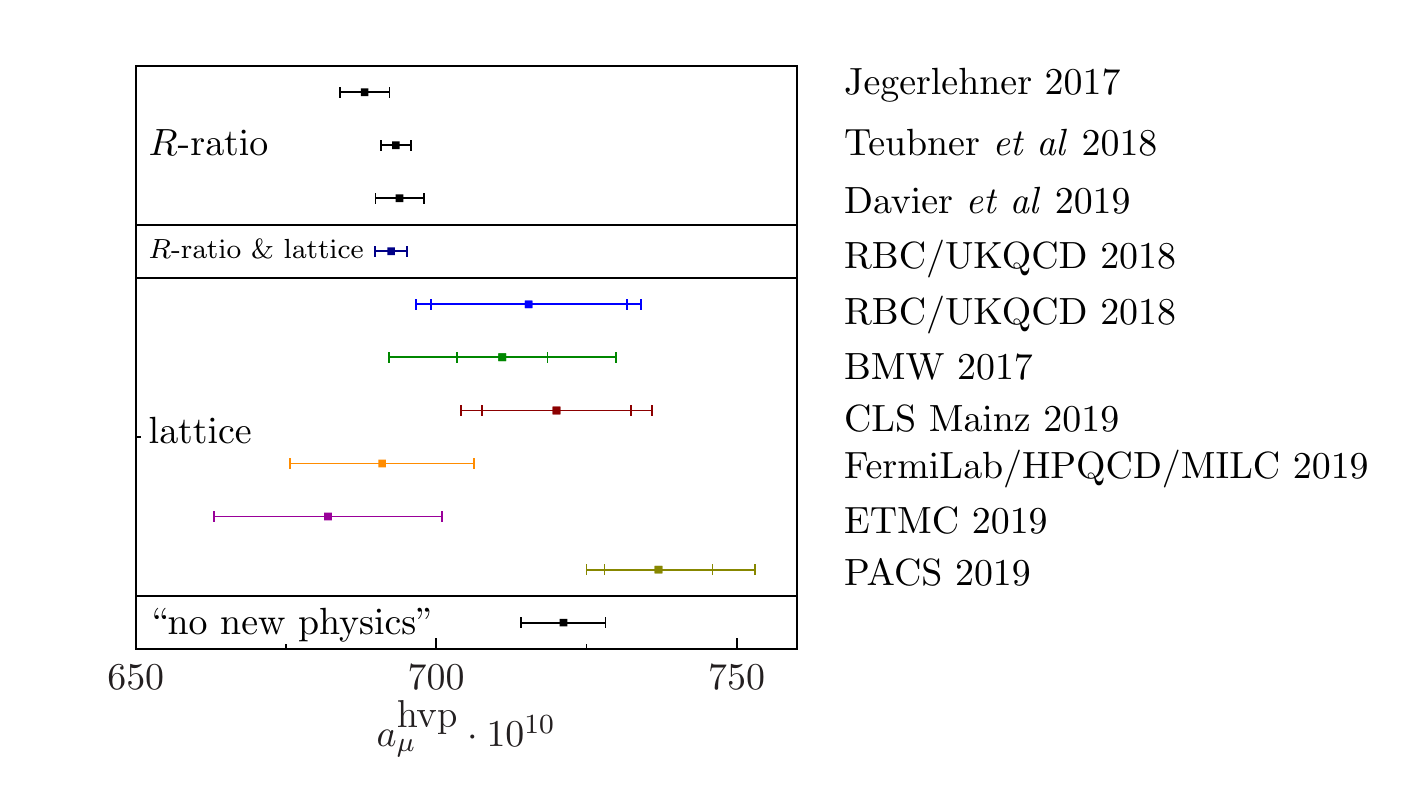}
 \vspace{-0.7cm}
 \caption{Comparison of various determinations of the HVP contribution to the anomalous magnetic 
moment of the muon.}
\label{fig:HVPcomp}
\end{figure}

Figure \ref{fig:HVPcomp} shows a comparison of results for the total HVP contribution to $a_\mu$. 
The upper panel shows determinations using the $R$-ratio data. The coloured points in the panel 
labeled ``lattice'' shows the published lattice results from various collaboration. The lowest 
point in the plot (``no new physics'') denotes the result when subtracting all other 
Standard Model contributions 
(as in table~\ref{tab:amucontr}) from the experimental result, i.e.\ the value that $\amuhvp$ would 
have to take for the Standard Model prediction to be in agreement with experiment. Clearly, at the 
current state-of-the-art, lattice calculations are not yet precise enough to distinguish between 
the $R$-ratio results and the ``no new physics'' scenario. \par
Furthermore, one can see that the 
smallest and largest lattice results disagree at a level of about $2\sigma$. Slight tensions 
between 
lattice results will have to be subject to further investigation in the future, in particular once 
the collaborations reduce the errors, to make sure to achieve consensus between the 
various lattice results. A possible approach is the comparison of more 
intermediate quantities, e.g. time-moments \cite{Chakraborty:2014mwa} of the vector correlator or 
$\amuhvp$ calculated from a time window~\cite{Blum:2018mom}. 
\par
Finally, the point in the second panel in figure \ref{fig:HVPcomp} shows a result from RBC/UKQCD 
\cite{Blum:2018mom} combining lattice and $R$-ratio results using a window method, where the vector 
correlator from small and large distances is taken from the $R$-ratio and intermediate distances 
from a lattice calculation. The point shown here uses the $R$-ratio data compilation from 
``Jegerlehner 2017'' \cite{alphaQED} and clearly shows that it is possible to improve the $R$-ratio 
results by 
supplementing it with data from a lattice calculation.

\par
\par
\begin{figure}[h]
 \centering 
 \includegraphics[scale=1.0]{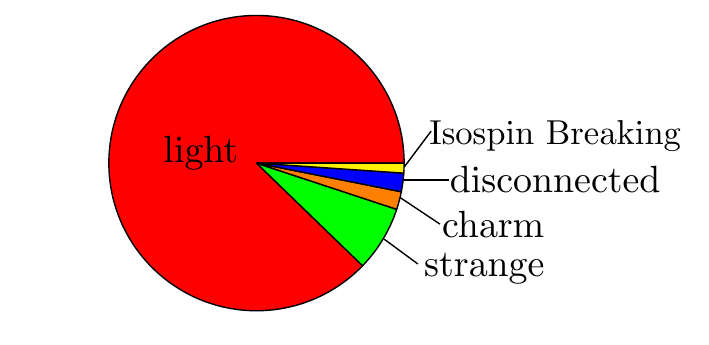}
 \includegraphics[scale=1.0]{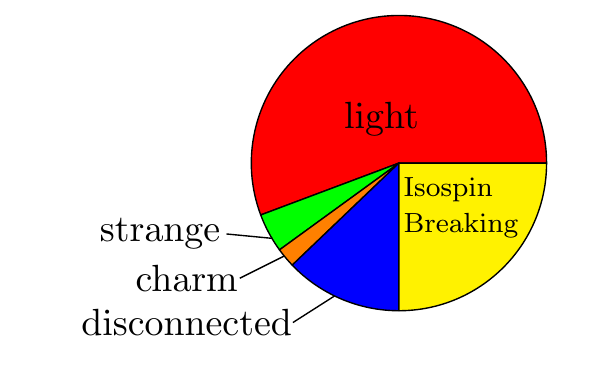}\\
\hspace{1cm}$a^\textnormal{\scriptsize HVP}_\mu$\hspace{7.8cm}$\delta a^\textnormal{\scriptsize 
HVP}_\mu$
 \caption{The relative size of the various flavour contributions to the HVP is shown on the left. 
The plot on the right shows the relative size of the average (statistical + systematic) error on 
the various flavour contributions from the available lattice results.}
\label{fig:HVPsize}
\end{figure}

The relative contribution of the various quark-flavours to the total HVP contribution from lattice 
calculations is shown by the pie chart on the left-hand side of figure \ref{fig:HVPsize}. Clearly, 
the by far biggest contribution comes for the light-quark connected diagram, followed by the 
strange-quark contribution. 

\par
The total errors on the lattice results for the HVP shown in figure \ref{fig:HVPcomp} are all of 
the order of $2-3\%$. This is clearly not yet competitive with the $R$-ratio results, which would 
require an accuracy of $\lesssim 1\%$, or even the required $0.2\%$ to match the precision of the 
upcoming experiments. The relative contribution to the error on the average lattice calculation is 
shown in the pie chart on the right-hand side of figure \ref{fig:HVPsize}. The 
error on the lattice results is dominated by the error on the light-quark 
contribution. The main challenges for reaching sub-percent precision for the light-quark 
contributions, that were discussed in 
previous sections, are controlling the statistical noise in the long-distance tail, having 
careful and reliable estimates of finite volume effects as well as a precise scale setting.
\par
The second biggest contribution to the error on the average lattice calculation (cf. right-hand 
side of figure \ref{fig:HVPsize}) comes from isospin breaking corrections (or the systematic error 
made by not including those effects). 
Given that the first calculations of isospin breaking corrections have only recently become 
available, progress and further reduction of statistical error is to be expected in the near 
future. However, it is also important to study the effects of including QED for the sea quarks, 
since these contributions could potentially be important at the level of sub-percent precision for 
the hadronic vacuum polarisation. 
\par
For the future, the first goal is to obtain calculations of $\amuhvp$ from lattice calculations at 
a precision of $\lesssim 1\%$, at which the lattice becomes competitive to $R$-ratio determinations.
Given the recent progress presented at this conference, first results at $1\%$ precision could 
be available within the time frame of a year.
\par
Besides the anomalous magnetic moment of the muon, the hadronic vacuum polarisation also enters in
other quantities, like the running of the electromagnetic coupling and the running of the 
electroweak mixing angle. Progress in this direction was presented by Mainz at this conference~ 
\cite{Ce:2019imp}. 
\section{Hadronic Light-by-Light Scattering}
The hadronic light-by-light scattering contribution (cf. diagram on the right-hand side of 
figure~\ref{fig:hadrdiagrams}) enters the anomalous magnetic moment at order $\alpha^3$. The value 
that is 
often used for the Standard Model prediction is the so-called ``Glasgow-consensus'' 
\cite{Prades:2009tw}, which includes model-dependent estimates of various contributions to the 
light-by-light scattering, with the largest contribution coming from the $\pi^0$-pole. 
Recent work in progress on dispersion relations for the hadronic light-by-light scattering can 
be found, e.g., in \cite{Colangelo:2018mtw,Hoferichter:2018dmo,Pauk:2014rfa} and references 
therein. 
\par
In the following, I will discuss the progress of \textit{ab initio} calculations of the hadronic 
light-by-light scattering using lattice QCD. 
\subsection{Light-by-Light from the Lattice}
\begin{figure}[h]
\centering
 \includegraphics[width=0.3\textwidth]{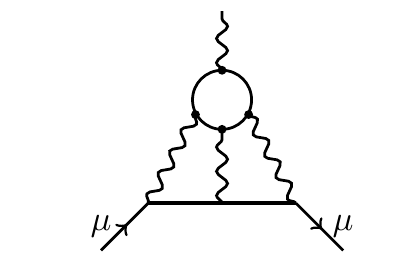}
 \hspace{1cm}
  \includegraphics[width=0.3\textwidth]{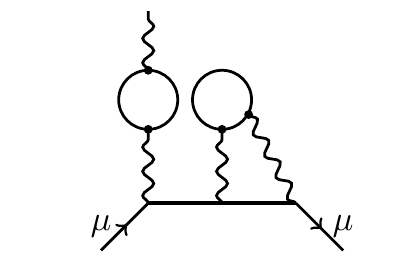}
   \vspace{-0.5cm}
 \caption{Fully quark-connected (left) and leading quark-disconnected (right) contributions to the 
hadronic light-by-light scattering.} 
\label{fig:lblwick}
\end{figure}

The hadronic part of the light-by-light scattering amplitude is written in terms of the expectation 
value of four electromagnetic currents
 \begin{equation}
  \Pi_{\mu\nu\lambda\rho}(q_1,q_2,q_3) = \int \textnormal{d}^4 x_1\textnormal{d}^4 
x_2\textnormal{d}^4 x_3 \,
e^{-i(q_1x_1+q_2x_2+q_3x_3)} 
\left<j^\gamma_\mu(x_1)j^\gamma_\nu(x_2)j^\gamma_\lambda(x_3)j^\gamma_\rho(0)\right>\,.
 \end{equation}
The fully quark-connected as well as the leading quark-disconnected Wick contraction are shown in 
figure \ref{fig:lblwick}. All other possible quark-disconnected contractions are $SU(3)$-flavour 
suppressed. 
\par
\par
Currently, two collaborations -- Mainz and RBC/UKQCD -- are actively working on calculating the 
hadronic light-by-light scattering contribution $\amulbl$ to the anomalous magnetic moment of the 
muon from the lattice. Both collaborations use position space approaches and their methods and 
status are described 
in sections \ref{subsec:lblmainz} and \ref{subsec:lblrbcukqcd}, respectively.

\subsection{Mainz Status}
\label{subsec:lblmainz}
In the position-space method developed by Mainz \cite{Green:2015mva,Asmussen:2016lse}
the hadronic light-by-light scattering contribution to $a_\mu$ is written as\\[-0.4cm]
 \begin{equation}
\amulbl = \frac{m_\mu e^6}{3} \int\!\textnormal{d} x^4\textnormal{d} y^4\, 
\overline{\mathcal{L}}_{[\rho,\sigma];\mu\nu\lambda}(x,y)\,\, i 
\hat{\Pi}_{\rho;\mu\nu\lambda\sigma}(x,y)\,.
\label{eq:mainzlbl}
 \end{equation}
 The hadronic part is given by the four-point function\\[-0.3cm]
 \begin{equation}
  \hat{\Pi}_{\rho;\mu\nu\lambda\sigma} = \int\!\textnormal{d} z^4\,\, iz_\rho 
\left<j^\gamma_\mu(x)\,j^\gamma_\nu(y)\,j^\gamma_\lambda(0)\,j^\gamma_\sigma(z)\right>
 \end{equation}
where $z$ is the position of the vertex with the external photon, and $x$, $y$ and $0$ the 
positions of the quark-photon vertices with the internal photon lines. The QED part of the 
light-by-light diagram is given by an electromagnetic 
kernel function $\overline{\mathcal{L}}_{[\rho,\sigma];\mu\nu\lambda}(x,y)$, that can be computed 
directly in the continuum and infinite volume limit. 
\par
\begin{figure}[h]
 \centering
 \includegraphics[width=0.47\textwidth]{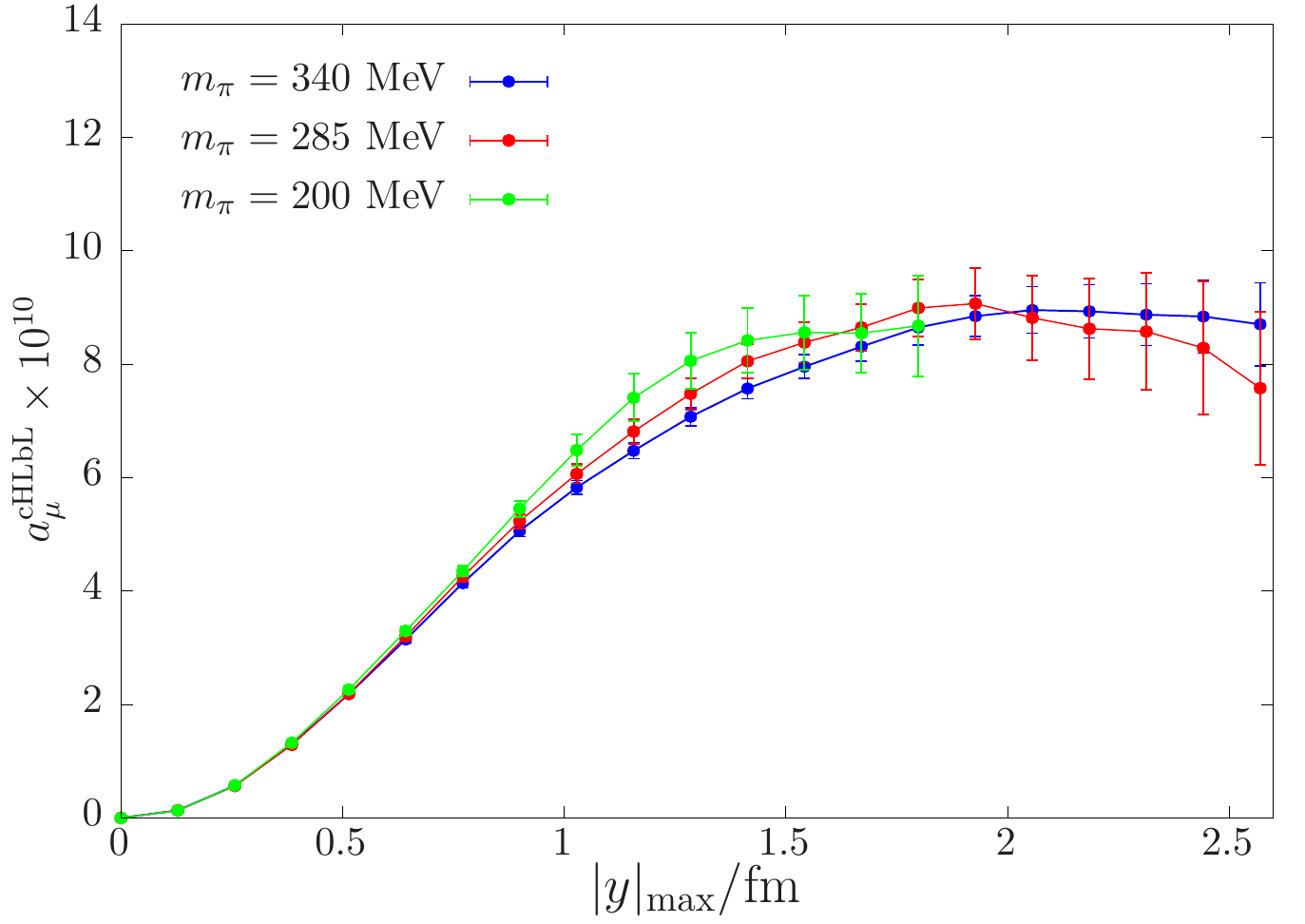}
 \caption{The quark-connected contribution to $\amulbl$ for different pion masses. The data has 
been 
integrated up to $|y|_\textnormal{\scriptsize max}$. Plot is taken from \cite{Asmussen:2019act}.}
 \label{fig:mainzlbl}
\end{figure}
\par
The current status of calculating $\amulbl$ using this method was presented 
at this conference \cite{Asmussen:2019act} and some results for the fully quark-connected 
contribution are shown in figure \ref{fig:mainzlbl}.
The plot in figure \ref{fig:mainzlbl} shows $\amulbl$ for three different pion masses at fixed 
lattice spacing. The data has been partially integrated up to $|y|_\textnormal{\scriptsize max}$, 
such that $\amulbl$ can be extracted from a plateau at large enough values of 
$|y|_\textnormal{\scriptsize max}$.

\subsection{RBC/UKQCD Status}
\label{subsec:lblrbcukqcd}
In the position space approach proposed by RBC/UKQCD \cite{Blum:2015gfa} the full hadronic 
light-by-light scattering diagram including the muon and photon propagators is treated on the 
lattice. To make this calculation feasible, position space sampling is used, where the sum 
over the position of two of the quark-photon vertices is sampled stochastically. For the photon 
propagators RBC/UKQCD uses either the QED$_L$ \cite{Hayakawa:2008an} prescription leading to 
power-law finite volume corrections or the infinite volume formulation (QED$_\infty$) of the 
photon propagator as proposed in \cite{Blum:2017cer}. 
\par
\begin{figure}
\centering
\hspace{-1.3cm}
 \includegraphics[height=4.6cm]{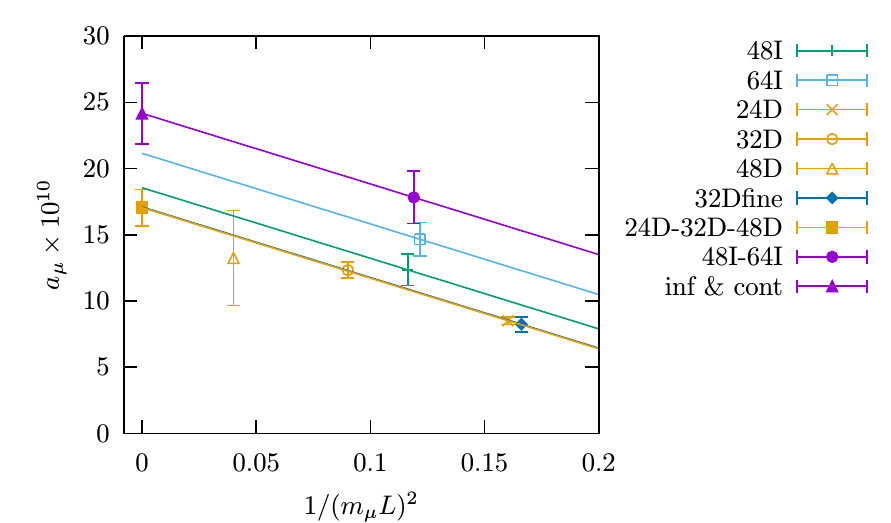}
\hspace{0.3cm}
\includegraphics[height=4.4cm]{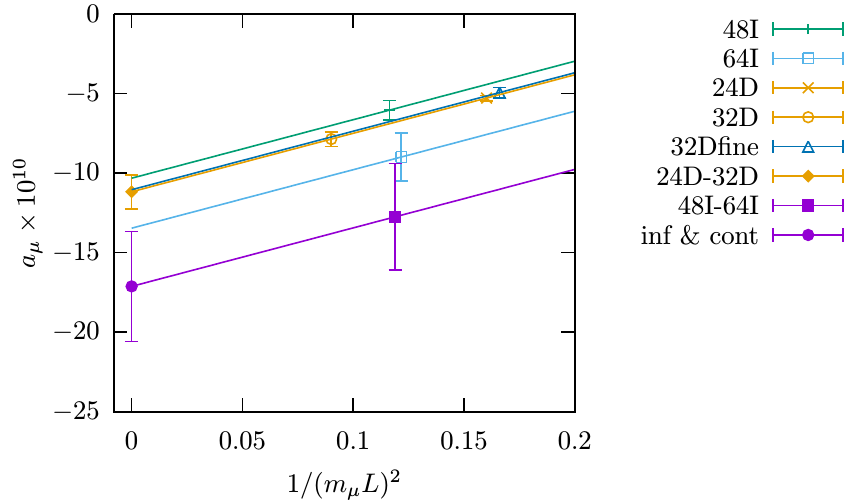}
\vspace{-0.2cm}
\caption{Infinite volume and continuum extrapolation of $\amulbl$ using QED$_L$ for the photons for 
the fully connected diagram (left) and the leading disconnected diagram (right). In both plots the 
purple point at $1/(m_\mu L)^2=0$ shows the result extrapolated 
to infinite volume and continuum. The plots are taken from~\cite{Blum:2019ugy}.}
\label{fig:lblrbcukqcd}
\end{figure}

Figure \ref{fig:lblrbcukqcd} shows results from the recent paper \cite{Blum:2019ugy} using QED$_L$ 
for the photon propagator. $\amulbl$ obtained on various different gauge ensembles is plotted 
against $1/(m_\mu L)^2$ with the box size $L$ for the fully quark-connected diagram (left) and the 
leading quark-disconnected diagram (right). The lines in the 
plots show the infinite volume and continuum extrapolation with the purple point at $1/(m_\mu 
L)^2=0$ showing the result extrapolated to the infinite volume and continuum limit.
\par
The quark-connected and leading quark-disconnected contribution are found to have opposite signs, 
and the result for the sum of both contribution is \cite{Blum:2019ugy}\\[-0.3cm]
\begin{equation}
\amulbl=7.20(3.98)(1.65)\times 10^{-10}\,. 
\label{eq:amlblresult}
\end{equation}
extrapolated to the physical point. The result (\ref{eq:amlblresult}) is consistent with the model 
estimates used in the Glasgow Concensus (cf. table \ref{tab:amucontr}), albeit with larger 
uncertainties. 
\par
\begin{figure}[h]
 \centering
 \includegraphics[width=0.27\textwidth]{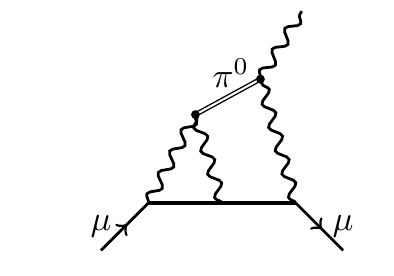}
 \vspace{-0.5cm}
 \caption{$\pi^0$ pole contribution to the hadronic light-by-light scattering}
 \label{fig:pi0pole}
\end{figure}

In addition RBC/UKQCD presented progress \cite{BlumLat19,JinLat19} on calculating $\amulbl$ 
using QED$_\infty$ at this conference. The long-distance tail of the hadronic light-by-light 
scattering is dominated by the $\pi^0$-pole contribution (cf. diagram in figure \ref{fig:pi0pole}). 
This can be used to further improve the calculation of $\amulbl$ by supplementing the 
long distance by the pion-pole contribution calculated either from a model or directly from the 
lattice using the pion transition form factor $\pi^0\rightarrow\gamma\gamma$ and work in progress 
in that direction was presented \cite{BlumLat19,JinLat19}. 

\subsection{Summary}
Currently two collaborations are actively working on determining the hadronic light-by-light 
scattering contribution to $a_\mu$ from first principles using lattice calculations. RBC/UKQCD has 
recently published \cite{Blum:2019ugy} the first result extrapolated to the continuum and infinite 
volume limit and Mainz has presented \cite{Asmussen:2019act} promising work in progress using their 
position space approach \cite{Green:2015mva,Asmussen:2016lse}.
\par
For the hadronic light-by-light scattering contribution to explain the discrepancy between 
experimental measurement and Standard Model prediction of $a_\mu$, one would need a value which is 
about three times larger than the number quoted in the Glasgow consensus. However, current lattice 
calculations suggest, that this is very unlikely. \par
To match the precision of the upcoming experimental results for $g-2$, the hadronic light-by-light 
scattering amplitude needs to be determined to a precision of about $10\%$. A promising proposal to 
reduce the statistical noise from the long-distance contribution is to constrain lattice data at 
long distances by the pion-pole contribution. A lattice calculation of the pion-pole 
contribution requires the 
pion transition form factor $\pi^0\rightarrow\gamma\gamma$ (see 
\cite{Gerardin:2016cqj,Gerardin:2019vio} for recent calculations from the Mainz collaboration).

\section{Final Remarks}
The persistent discrepancy between the Standard Model prediction (table \ref{tab:amucontr}) and the 
experimental result (equation (\ref{eq:amuexp})) for the anomalous magnetic moment of the muon has 
triggered a tremendous effort within the lattice community to calculate the hadronic contributions 
to $a_\mu$ from first principles. The upcoming experiments at Fermilab \cite{Venanzoni:2014ixa} and 
JPARC \cite{Otani:2015jra} aim to further reduce the uncertainty of the experimental result by a 
factor of $4$. To match the precision of these upcoming experiments one finally 
has to determine the hadronic vacuum polarisation contribution $\amuhvp$ to a precision of about 
$0.2\%$ and the hadronic light-by-light scattering $\amulbl$ to about $10\%$ accuracy.
\par
In terms of the hadronic vacuum polarisation, which is the leading hadronic contribution to 
$a_\mu$, several results from different collaborations with a precision of $2-3\%$ are available 
and summarised in figure \ref{fig:HVPcomp}. Recent progress presented at this conference suggests, 
that the first results at around $1\%$ precision could be available within the time frame of a 
year.

\section*{Acknowledgments}
The author thanks N. Asmussen, C. Aubin, T. Blum, C. DeTar, D. Giusti, C. Lehner, L. Lellouch, A. 
Meyer, B. Toth and G. von Hippel for useful discussions and/or providing material 
prior to the conference. \par 
VG has received funding from the European Research Council (ERC) 
under 
the 
European Union's Horizon 2020 research and innovation programme under grant agreement No 757646.

%\newpage

\end{document}